\documentclass[twocolumn,unsortedaddress,superscriptaddress, nofootinbib]{revtex4}
\usepackage{graphicx}
\usepackage{color}
\usepackage{amsmath}
\usepackage{ulem} 
\usepackage{bm} 
\usepackage{comment}
\usepackage[breaklinks=true,colorlinks,citecolor=blue,linkcolor=blue,urlcolor=blue]{hyperref}

\newcommand{\be}{\begin{equation}}
\newcommand{\ee}{\end{equation}}
\newcommand{\bea}{\begin{eqnarray}}
\newcommand{\eea}{\end{eqnarray}}

\begin{document}

\title{Out-of-plane heat transfer in van der Waals stacks through electron-hyperbolic phonon coupling}

\author{Klaas-Jan Tielrooij\normalfont\textsuperscript{$\ddagger$,}} \email{Correspondence: klaas-jan.tielrooij@icfo.eu, frank.koppens@icfo.eu} 
\author{Niels C.\ H.\ Hesp\normalfont\textsuperscript{$\ddagger$,}}  \affiliation{ICFO - Institut de Ci\`{e}ncies Fot\`{o}niques, The Barcelona Institute of Science and Technology, Castelldefels (Barcelona) 08860, Spain} 
\author{Alessandro Principi} \affiliation{Radboud University, Institute for Molecules and Materials, NL-6525 AJ Nijmegen, the Netherlands}
\author{Mark B.\ Lundeberg} \affiliation{ICFO - Institut de Ci\`{e}ncies Fot\`{o}niques, The Barcelona Institute of Science and Technology, Castelldefels (Barcelona) 08860, Spain}
\author{Eva A.\ A.\ Pogna} \affiliation{IFN-CNR, Dipartimento di Fisica, Politecnico di Milano, Piazza L. da Vinci 32, 20133 Milano, Italy}
\author{Luca Banszerus} \affiliation{JARA-FIT and 2nd Institute of Physics, RWTH Aachen University, 52074 Aachen, Germany}
\author{Zolt\'{a}n Mics} \affiliation{Max Planck Institute for Polymer Research, Ackermannweg 10, Mainz 55128, Germany}
\author{Mathieu Massicotte}
\author{Peter Schmidt}
\author{Diana Davydovskaya}  \affiliation{ICFO - Institut de Ci\`{e}ncies Fot\`{o}niques, The Barcelona Institute of Science and Technology, Castelldefels (Barcelona) 08860, Spain}
\author{David G.\ Purdie}
\author{Ilya Goykhman}
\author{Giancarlo Soavi}
\author{Antonio Lombardo}
\affiliation{Cambridge Graphene Centre, University of Cambridge, Cambridge CB3 OFA, UK}
\author{Kenji Watanabe}
\author{Takashi Taniguchi} \affiliation{National Institute for Material Science, 1-1 Namiki, Tsukuba 305-0044, Japan}
\author{Mischa Bonn}
\author{Dmitry Turchinovich} \affiliation{Max Planck Institute for Polymer Research, Ackermannweg 10, Mainz 55128, Germany}
\author{Christoph Stampfer} \affiliation{JARA-FIT and 2nd Institute of Physics, RWTH Aachen University, 52074 Aachen, Germany}
\author{Andrea C.\ Ferrari} 
\affiliation{Cambridge Graphene Centre, University of Cambridge, Cambridge CB3 OFA, UK}
\author{Giulio Cerullo} \affiliation{IFN-CNR, Dipartimento di Fisica, Politecnico di Milano, Piazza L. da Vinci 32, 20133 Milano, Italy}
\author{Marco Polini} \affiliation{Istituto Italiano di Tecnologia, Graphene Labs, Via Morego 30, I-16163 Genova, Italy}
\author{Frank H.\ L.\ Koppens} \email{Correspondence: klaas-jan.tielrooij@icfo.eu, frank.koppens@icfo.eu}  \affiliation{ICFO - Institut de Ci\`{e}ncies Fot\`{o}niques, The Barcelona Institute of Science and Technology, Castelldefels (Barcelona) 08860, Spain} \affiliation{ICREA - Instituci\'o Catalana de Re\c{c}erca i Estudis Avancats, 08010 Barcelona, Spain}

\begin{abstract}
\textbf{
Van der Waals heterostructures have emerged as promising building blocks that offer access to new physics, novel device functionalities, and superior electrical and optoelectronic properties \cite{Geim2013, Dean2010, Mayorov2011, Britnell2012, Wang2013, Dean2013, Nanoscale15}. Applications such as thermal management, photodetection, light emission, data communication, high-speed electronics and light harvesting \cite{Balandin2008, Britnell2013, Koppens2014, Lopez2014, Bonaccorso2015, Mics2015, Kim2015, Massicotte2016, BonaNP10} require a thorough understanding of (nanoscale) heat flow. Here, using time-resolved photocurrent measurements we identify an efficient out-of-plane energy transfer channel, where charge carriers in graphene couple to hyperbolic phonon polaritons \cite{Caldwell2014, Dai2014, Basov2016} in the encapsulating layered material. This hyperbolic cooling is particularly efficient, giving picosecond cooling times, for hexagonal BN, where the high-momentum hyperbolic phonon polaritons enable efficient near-field energy transfer. We study this heat transfer mechanism through distinct control knobs to vary carrier density and lattice temperature, and find excellent agreement with theory without any adjustable parameters. These insights may lead to the ability to control heat flow in van der Waals heterostructures.}
\end{abstract}

\maketitle

Owing to its large in-plane thermal conductivity, graphene has been suggested as material for the thermal management of nanoscale devices \cite{Balandin2008}. At the same time, graphene is well-known for its ability to convert incident light into electrical heat, i.e. hot electrons that can be used to generate photocurrent, with applications in photodetection, data communication and light harvesting \cite{Song2011, Gabor2011, Koppens2014}. Understanding, and ultimately controlling, heat flow in graphene-van der Waals heterostructures is therefore of paramount importance. For example, a short cooling time of graphene hot carriers is advantageous for thermal management and for high switching rates of photodetectors (PDs) for data communication, whereas a long cooling time is favorable for photodetection sensitivity \cite{Song2011, Gabor2011, Koppens2014}. Of particular relevance are heterostructure devices that contain high-quality graphene encapsulated by layered materials, such as hexagonal BN (hBN) and MoS$_2$, which have the potential to crucially improve the performance of electronic and optoelectronic devices \cite{Dean2010,Geim2013}. \let\thefootnote\relax\footnote{$\ddagger$ Equal contribution.}
\

\begin{figure} [h!!!!!]
   \centering
  \includegraphics [scale=1]
   {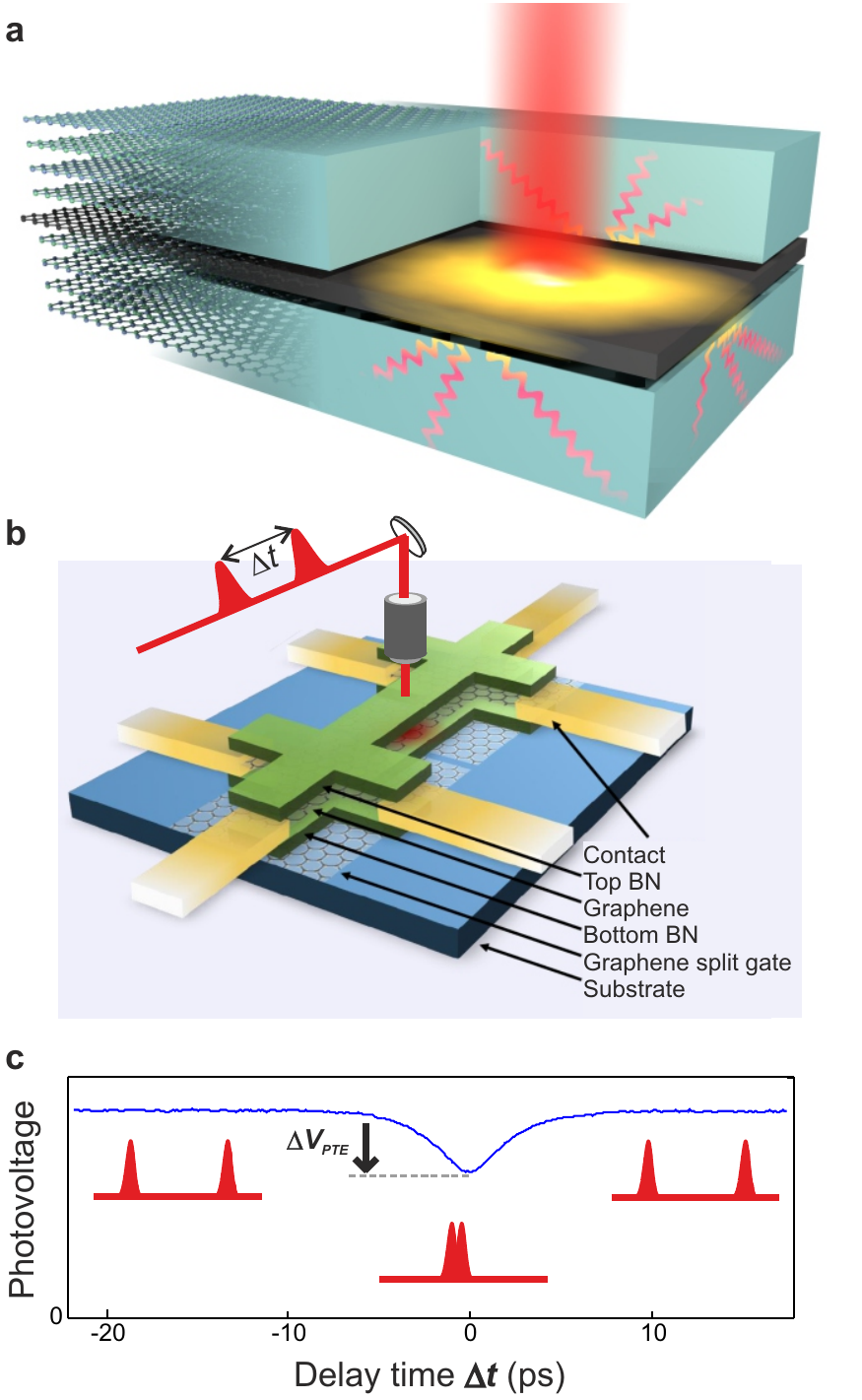}
      \caption{ \textbf{Hot carrier cooling in hBN-encapsulated graphene.} \textbf{a)} Schematic representation of out-of-plane heat transfer in hBN-encapsulated graphene: highly efficient near-field emission from graphene hot carriers into directional hyperbolic phonon polaritons of the encapsulating material. \textbf{b)} Schematic drawing of the hBN-encapsulated graphene device with graphene split gates. By applying a different voltage to the two split gates we create a $pn$-junction in the encapsulated graphene layer, which we illuminate with two ultrafast pulses that arrive with a variable time delay $\Delta t$. \textbf{c)} The resultant photo-thermoelectric photovoltage measured between two contacts shows a dip when the two ultrafast pulses overlap. This is due to the nonlinear relation between hot carrier temperature and incident light intensity. The dynamics of the photovoltage dip $\Delta V_{\rm PTE}$ as a function of time delay $\Delta t$ therefore correspond to the cooling dynamics of the hot carriers at the $pn$-junction. }
\label{Fig1}
   \end{figure}

A number of cooling pathways for graphene carriers have been proposed, involving among others strongly coupled optical phonons \cite{Kampfrath2005,Mihnev2016,Brida2013}, acoustic phonons \cite{Bistritzer2009, Song2012, Graham2013, Betz2013}, substrate phonons \cite{Low2012} and plasmons \cite{Hamm2016} (see also Appendix 1). Here, using several experimental approaches, we show that cooling in graphene encapsulated by hBN is governed by out-of-plane coupling of graphene electrons to special polar phonon modes that occur in layered materials (LMs): hyperbolic phonon polaritons, where $\epsilon_{xx}\epsilon_{zz} <$ 0, with $\epsilon_{xx}$ and $\epsilon_{zz}$  the permittivity parallel and perpendicular to the LM plane. Owing to this property, these materials carry deep sub-wavelength, ray-like optical phonon polaritons. For hBN, within the two Reststrahlen bands a large number of hyperbolic phonon modes exist with high momenta, far outside the light cone. The most notable modes occur at an energy of $\sim$100 meV and $\sim$180 meV \cite{Caldwell2014}, such that energy overlap with the graphene hot-carrier distribution is substantial. The unusual hyperbolic character gives rise to a very high density of optical states, and thus large thermal energy densities \cite{Caldwell2014, Dai2014}, thereby providing a potentially efficient cooling pathway for hot carriers in graphene. By near-field coupling between graphene and hBN, efficient energy transfer from hot carriers to hyperbolic phonon polaritons is possible \cite{Principi2016}. Here we show that the measured carrier dynamics of hBN-encapsulated graphene can be explained by this hyperbolic cooling process, as illustrated in \textcolor{blue}{\textbf{\ref{Fig1}a}}.
\

We use ultrafast time-resolved photocurrent measurements and show the results obtained from one hBN-encapsulated, Hall-bar shaped, exfoliated graphene device (see \textcolor{blue}{\textbf{\ref{Fig1}b}}). The device contains chemical vapor deposited (CVD) graphene split gates underneath the bottom hBN (thickness: 70 nm), in order to generate a $pn$-junction in the middle of the device. A second hBN-encapsulated device with metallic split gates gave fully consistent results. With 800 nm light incident on the $pn$-junction, a photovoltage is generated by the photo-thermoelectric (PTE) effect \cite{Song2011, Gabor2011} (see the photocurrent and reflection data in \ref{SuppFigSpatialImage} and the characteristic PTE sixfold pattern in \ref{SuppFigSixfold}). By varying the delay $\Delta t$ between two sub-picosecond pulses we extract the carrier dynamics from the photovoltage signal $\Delta V_{\rm PTE} (\Delta t)$ (see \textcolor{blue}{\textbf{\ref{Fig1}c}}). Since the photo-thermoelectric voltage scales with the light-induced increase in carrier temperature, the decay dynamics of $\Delta V_{\rm PTE} (\Delta t)$ closely mimic the cooling dynamics of the hot electron system (see \ref{SuppFigNonlinearArtefact} for an analysis of the extent to which the experimentally observed decay dynamics correspond to the underlying cooling dynamics). In the case of exponential decay dynamics, which we observe above a lattice temperature of $\sim$200 K, we extract an experimental cooling time scale $\tau_{\rm exp}$ by describing the decay dynamics with $\Delta V_{\rm PTE} (\Delta t) \propto e^{-\Delta t/\tau_{\rm exp}}$.
\

\begin{figure} [h!!!!!]
   \centering
  \includegraphics [scale=0.79]
   {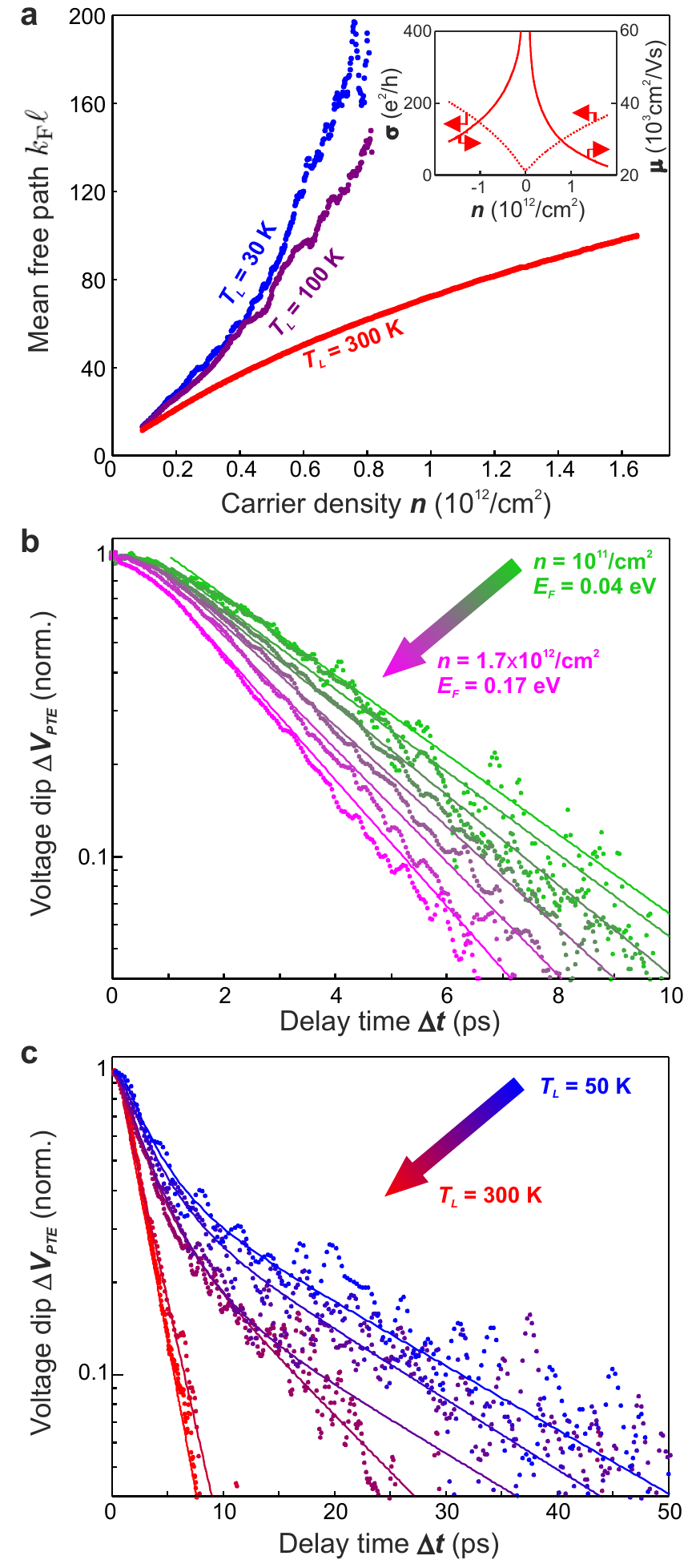}
      \caption{ \textbf{The effect of doping and lattice temperature.} \textbf{a)} From four-probe transport measurements, we extract the dimensionless mean free path parameter $k_{\rm F} \ell$  as a function of carrier density for three different lattice temperatures. These data show that carrier scattering in our hBN-encapsulated graphene is much less efficient than in SiO$_2$-supported graphene where $k_{\rm F} \ell$ is typically 10 or less, and where disorder-assisted supercollisions constitute the dominant cooling mechanism \cite{Graham2013,Song2012,Betz2013}. The inset shows the conductivity and mobility vs.\ gate-induced carrier density. By comparing four-probe and two-probe measurements we find that each contact has a contact resistance of 5 --8 k$\Omega$. \textbf{b-c)} The dynamics of the photovoltage dip $\Delta V_{\rm PTE}$, showing that hot-carrier cooling becomes faster upon increasing the carrier density \textbf{(b)} and upon increasing the lattice temperature \textbf{(c)}. Panel \textbf{b} is measured at a lattice temperature of $T_L$ = 300 K and panel \textbf{c} at a carrier density of $n \approx 1.7 \times 10^{12}$/cm$^2$. The solid lines are bi-exponential guides to the eye. }
\label{Fig2}
   \end{figure}

First, we electrically characterize our device using four-probe measurements (see \textcolor{blue}{\textbf{\ref{Fig2}a}}) and find a mean free path of $k_{\rm F} \ell$ = 80--100 at a carrier density of $n = 1.7\times 10^{12}$/cm$^2$, corresponding to a mobility of 25,000--30,000 cm$^2$/Vs and momentum scattering time of 340--440 fs. As expected, this is much higher than SiO$_2$-supported devices with typically $k_{\rm F} \ell <$ 10, mobility $<$ 5,000 cm$^2$/Vs and momentum scattering time $<$ 100 fs (see e.g.\ Ref.\ \cite{Graham2013}). In such devices, carrier cooling is typically ascribed to disorder-assisted cooling to graphene acoustic phonons \cite{Song2012, Graham2013, Betz2013}.
\

To study hot-carrier cooling in our high-mobility, encapsulated devices we examine $\Delta V_{\rm PTE} (\Delta t)$ while varying graphene's most characteristic parameter, the carrier density $n$. In particular, we apply a gate voltage $V_{\rm L} = +V - V_{\rm D}$ to the left split gate and $V_{\rm R} = -V - V_{\rm D}$ to the right gate, such that there is always either a $pn$-junction or an $np$-junction, with equal electron and hole densities in the two graphene regions ($V_{\rm D}$ is the gate voltage that corresponds to the Dirac point). The incident laser fluence is typically 5 -- 40 $\mu$J/cm$^2$. The data show that cooling becomes faster upon increasing the carrier density (see \textcolor{blue}{\textbf{\ref{Fig2}b}}). We also vary the lattice temperature $T_{\rm L}$ and observe faster decay for increasing lattice temperature (see \textcolor{blue}{\textbf{\ref{Fig2}c}}, taken at $n \approx$ 10$^{12}$/cm$^2$). At low temperatures (below $\sim$200 K), cooling is non-exponential, whereas at room temperature we observe exponential decay of the photovoltage dip, with a timescale of $\tau_{\rm exp} \approx$ 2.5 ps (for $n = 1.7 \times 10^{12}$/cm$^2$). We independently verify this cooling time using two alternative measurement techniques that are both sensitive to electron cooling dynamics in different ways (see \ref{SuppFigPumpProbe}). Firstly, using ultrafast optical pump -- optical probe spectroscopy, which probes interband transitions \cite{Brida2013}, we find $\tau_{\rm exp} $= 2.55 ps for the decay of the absorption photobleaching. Secondly, using optical pump -- terahertz probe spectroscopy, which probes intraband transitions \cite{Mics2015}, we obtain $\tau_{\rm exp}$ = 2.2 ps for the decay of the photoconductivity. For these two experiments we use similar excitation conditions (an incident pulse fluence of 8--20 $\mu$J/cm$^2$) as in the photocurrent measurements, and the measurements are performed on two separate devices consisting of large-area, high-quality hBN-encapsulated CVD graphene as in Ref.\ \cite{Banszerus2015}. Thus, all three techniques consistently yield similar cooling times for hBN-encapsulated graphene.
\

We compare these observations with different cooling mechanisms, and note that the data are qualitatively and quantitatively inconsistent with in-plane cooling by scattering with graphene acoustic phonons through normal \cite{Bistritzer2009} or disorder-assisted collisions \cite{Song2012, Graham2013, Betz2013} (see Appendix 1 and \ref{SuppFigSupercollision}). In-plane cooling can also occur by scattering to graphene optical phonons, typically occurring on a sub-picosecond timescale \cite{Kampfrath2005,Brida2013}. Reference \cite{Mihnev2016} reports that for \textit{non-encapsulated} graphene this channel gives picosecond decay dynamics of the THz photoconductivity. However these results deviate in several ways from our observations for hBN-encapsulated graphene (see Appendix 1). For example, we observe a twofold increase in cooling time by decreasing $T_{\rm L}$ from 300 K to 200 K, whereas optical phonon cooling gives cooling dynamics that are independent of substrate temperature (for $E_{\rm F}$ = 0.3 eV), see Ref.\ \cite{Mihnev2016}. This indicates that, besides a fraction of hot graphene electrons possibly cooling through optical phonons, a different cooling mechanism plays an important role. Since we observe a striking effect of the hBN crystal slab thickness on the cooling dynamics (see \ref{SuppFigThinHBN} and Appendix 2), we propose an out-of-plane cooling mechanism where hot carriers in graphene lose their energy to remote polar phonons in the encapsulating layered material. The energy transfer can be understood from fluctuation electrodynamics, where any process that dissipates energy in the form of heat, has a reverse process that is driven by thermal fluctuations and thus becomes stronger at higher temperatures \cite{Nyquist1928, Mihnev2015}. An example of such paired processes is light absorption and blackbody radiation, which means that graphene emits thermal noise due to the dissipative real part (indicated by $\mathcal{R}$) of the frequency- and momentum-dependent optical sheet conductivity $\sigma (\omega, k)$. This thermal noise is efficiently absorbed by hBN, hBN being a lossy-polarizable material, leading to an energy transfer rate \cite{Principi2016}:
\be \mathcal{Q} = \iiint_{-\infty}^{\infty} \frac{d \omega d k_x d k_y}{ (2\pi)^3} [n_B(\omega, T_{\rm e}) - n_B(\omega, T_{\rm L}) ] M(\omega, k) \hspace{3mm}, \ee
\noindent where $n_B (\omega, T) = \frac{\hbar|\omega|}{e^{\hbar|\omega|/k_{\rm B}T} - 1}$ and $M (\omega, k) = 4 \frac{\mathcal{R}\{{Y (\omega, k)}\}\mathcal{R}\{{\sigma (\omega, k)}\}}{|Y (\omega, k) + \sigma (\omega, k)|^2}$ ($k_{\rm B}$ is Boltzmann's constant). Heat transfer from hot graphene electrons to hBN hyperbolic phonon polaritons is thus governed by the Bose factor $[n_B(\omega, T_{\rm e}) - n_B(\omega, T_{\rm L}) ]$, which describes the energy disequilibrium between hot graphene carriers and the cold hBN phonon system, and the impedance matching function $M (\omega, k)$, which is nonzero when the surface admittance $Y (\omega, k)$ has a nonzero real part. We calculate $Y (\omega, k)$ as in Ref.\ \cite{Tomadin2015} (see Methods for details) and find that nearby the hyperbolic hBN phonon frequencies $Y (\omega, k)$ is real over a large $k$-space area and relatively wide frequency band. Since $\sigma (\omega, k)$, calculated using the Random Phase Approximation \cite{Giuliani2005}, also has a significant real part, this leads to an impedance matching function $M$ approaching unity. Due to this near-field coupling to hyperbolic modes, the heat conductivity exceeds Planck's law for blackbody radiation by orders of magnitude (see \ref{SuppFigHeatConductivity}). The reason for this is that in vacuum, the $k$-space for blackbody radiation is limited to $k < \omega/c$ (with $c$ the speed of light), whereas this restriction is lifted in the near-field interaction with hBN hyperbolic phonon polaritons. This super-Planckian coupling to hyperbolic hBN phonons thus provides a highly efficient cooling channel for hot carriers in graphene. Cooling to hyperbolic modes also occurs in materials such as MoS$_2$, although there it is not as efficient as for hBN (see \ref{SuppFigHeatConductivity}).
\

\begin{figure} [h!!!!!]
   \centering
 \includegraphics [scale=0.7]
   {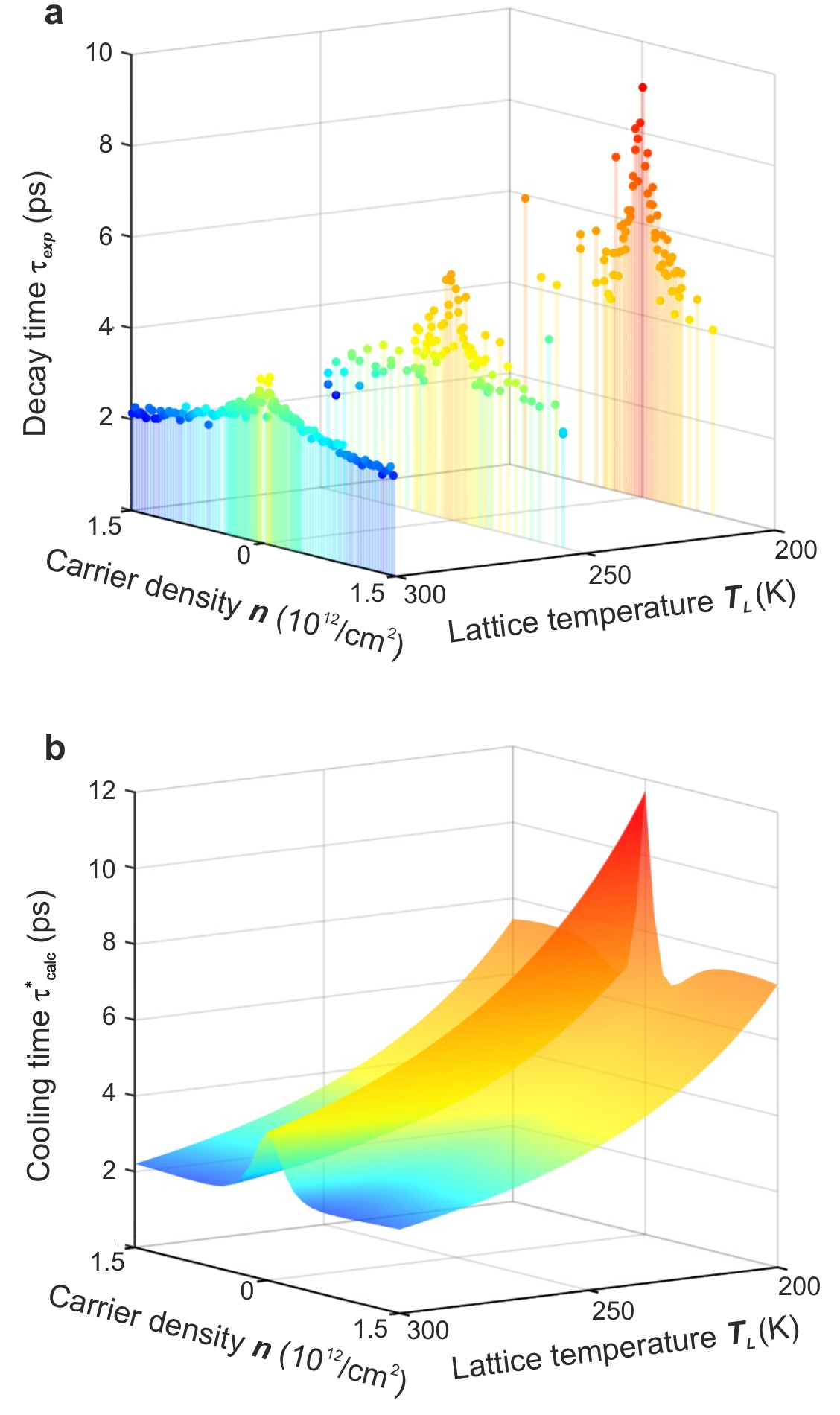}
      \caption{ \textbf{Qualitative comparison with hyperbolic hBN cooling.} Comparison between the experimental decay time $\tau_{\rm exp}$ extracted from the dynamics of the photovoltage dip \textbf{(a)} and the predicted near-equilibrium cooling time $\tau_{\rm calc}^*$ for super-Planckian cooling to hyperbolic hBN phonons \textbf{(b)}. Both in experiment and theory, we vary the carrier density and lattice temperature. }
\label{Fig3}
   \end{figure}

To compare our hyperbolic cooling theory with the experimental data, we examine the calculated energy transfer rate $\mathcal{Q}$. Here, solving Eq.\ (1) gives cooling dynamics with a cooling time scale \be \tau_{\rm calc} (T_{\rm e}, T_{\rm L}) = C_{\rm n} \frac{T_{\rm e} - T_{\rm L}}{\mathcal{Q}} \hspace{3mm} ,\ee
\noindent where $C_n$ is the electronic heat capacity of graphene at constant $n$ \cite{Principi2016}. In the limit of weak heating, where $T_{\rm e}$ approaches $T_{\rm L}$, we obtain exponential decay with the near-equilibrium timescale $\tau_{\rm calc} (T_{\rm e} \rightarrow T_{\rm L}) = \tau_{\rm calc}^* = C_n \big ( \frac{\partial \mathcal{Q}}{\partial T_{\rm e}} \bigl \rvert_{T_{\rm e}=T_{\rm L}} \big )^{-1}$, where $\frac{\partial \mathcal{Q}}{\partial T_{\rm e}}\bigl \rvert_{T_{\rm e}=T_{\rm L}}$ is the interfacial heat conductivity $\Gamma$. We compare the calculated near-equilibrium cooling time $\tau_{\rm calc}^*$ with the measured exponential decay time $\tau_{\rm exp}$ (see \textcolor{blue}{\textbf{\ref{Fig3}}}), although strictly speaking these experimental timescales correspond to the strong heating regime, where $T_{\rm e} \gg T_{\rm L}$. The reason for this is that our technique is not sensitive enough at low incident powers, which means that experimentally we cannot directly access the near-equilibrium cooling time $\tau_{\rm calc}^*$. Nevertheless, we find that our hyperbolic hBN cooling model semi-quantitatively reproduces the experimentally observed trends for the entire range of investigated carrier densities (up to 1.6$\times10^{12}$/cm$^2$) and lattice temperatures (200 -- 300 K). In particular, cooling slows down for lower lattice temperatures, which we attribute mainly to the smaller Bose factor in Eq.\ (1) and thus smaller energy transfer rate $\mathcal{Q}$. The calculations also reproduce the observation of a longer cooling time around the Dirac point. This is the result of the energy transfer rate $\mathcal{Q}$ smoothly decreasing towards zero carrier density, while the electronic heat capacity $C_n$ flattens around the Dirac point towards its neutral graphene value (see \ref{SuppFigHeatConductivity}). At higher carrier densities ($n > 10^{12}$/cm$^2$) the increasing energy transfer rate is compensated by the increasing heat capacity, leading to a weak dependence of cooling time on carrier density. Cooling to hyperbolic modes was also observed in a noise thermometry study \cite{Yang2017}. We further note that the two distinct hyperbolic modes contribute almost equally to the overall cooling time, with the lower-energy mode slightly dominant (see \ref{SuppFigHeatConductivity}). This can be attributed to more energy overlap in the Bose factor. We note that the hyperbolic cooling model reproduces the observed slower decay dynamics for encapsulation with very thin hBN flakes (see \ref{SuppFigThinHBN}). In this case, cooling is slower because overall there is a lower density of hyperbolic modes to couple to.
\

\begin{figure} [hhhhhh!!!!!!!!]
   \centering
 \includegraphics [scale=0.79]
   {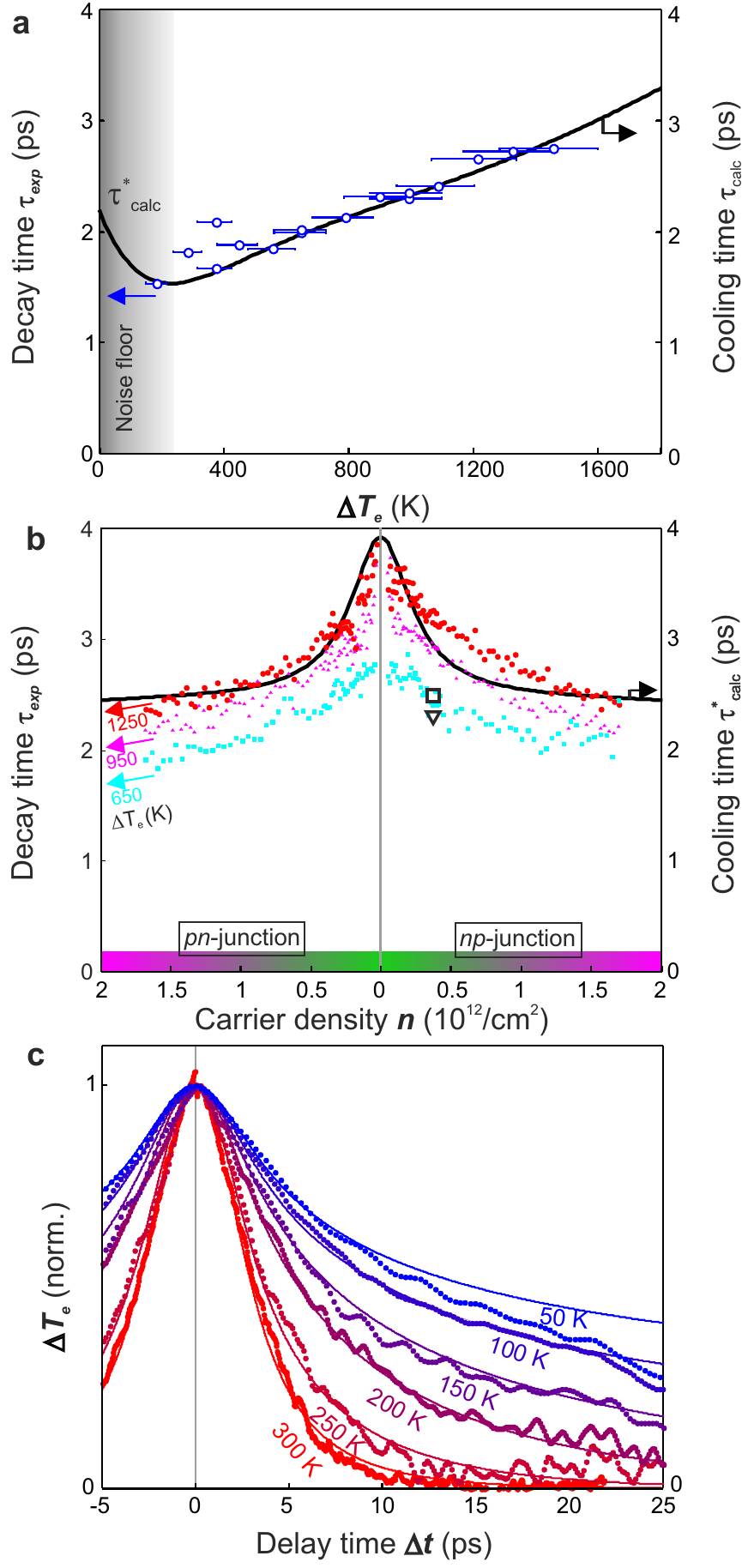}
      \caption{ \textbf{Quantitative comparison with hyperbolic hBN cooling. } 
\textbf{a)} Comparison between measured decay time $\tau_{\rm exp}$ (blue symbols and left vertical axis) and calculated cooling time $\tau_{\rm calc}$ (solid black line and right vertical axis) as a function of electron temperature \textit{without any adjustable parameters}. 
\textbf{b)} The measured decay time $\tau_{\rm exp}$ (symbols and left vertical axis) at room temperature as a function of carrier density for three different laser powers, compared with the near-equilibrium cooling time $\tau_{\rm calc}^*$ (solid line and right vertical axis) according to super-Planckian cooling to hyperbolic hBN phonons \textit{without any adjustable parameters}. The open square (triangle) shows the obtained decay time from time-resolved optical (terahertz) spectroscopy for $\Delta T_{\rm e} = 600 \pm 200$ K. 
\textbf{c)} Comparison of the complete cooling dynamics as measured (data points) and calculated (solid line) for $n = 0.06 \times 10^{12}$/cm$^2$ for varying lattice temperature. The calculated dynamics are convoluted with a Gaussian function representing the experimental time resolution. We find an initial temperature increase of $\Delta T =$ 300--500 K for all temperatures, except 1250 K for $T_{\rm L}$ = 300 K. The lower $\Delta T$ is related to the larger effective photocurrent spot size.}
\label{Fig4}
   \end{figure}

\begin{figure} [h!!!!!]
   \centering
  \includegraphics [scale=0.8]
   {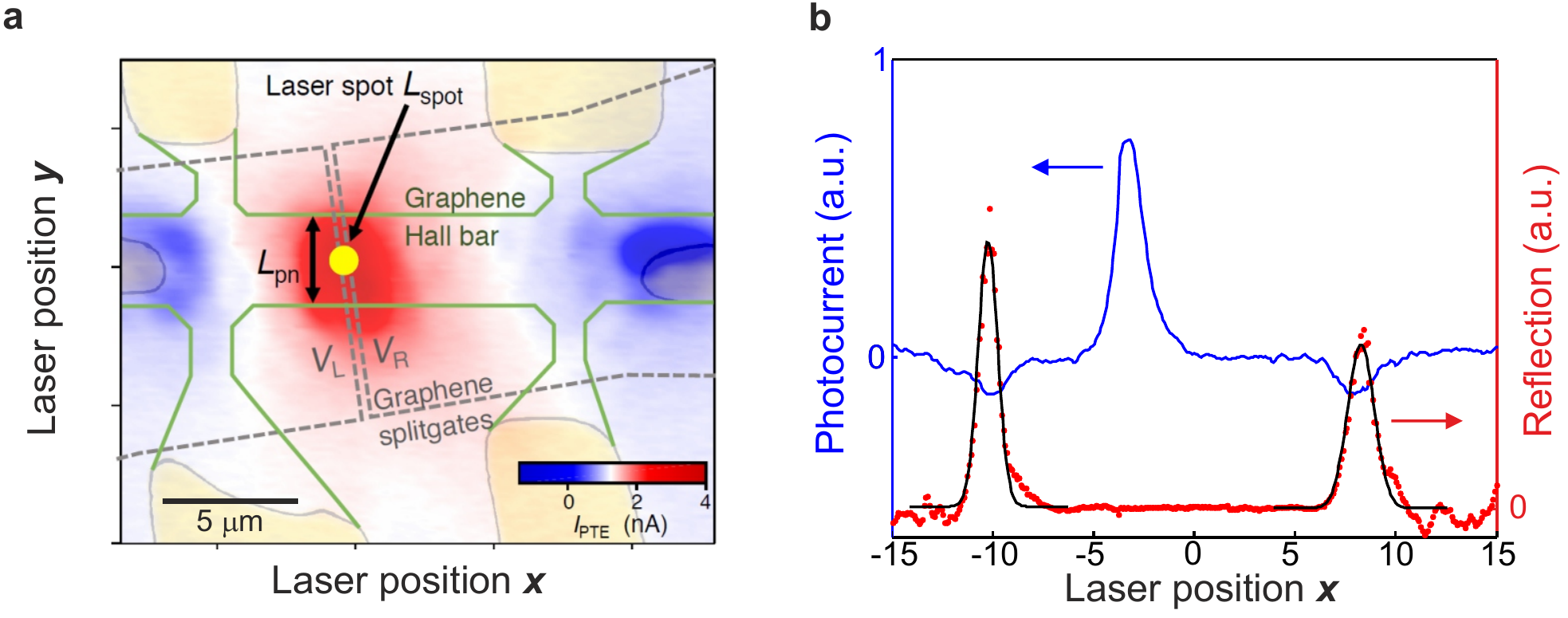}
\caption{ \textbf{Photocurrent and reflection data.}
      \textbf{a)} Spatial image, where the laser focus is scanned over the device, while photocurrent and laser reflection are simultaneously measured (at room temperature). The reflection image indicates the metal contacts (yellow) and we observe negative photocurrent at the metal-graphene interfaces and positive photocurrent at the $pn$-junction.  \textbf{b)} Photovoltage and spatial-derivative reflection line traces as indicated in panel \textbf{b}, with Gaussian fits to the reflection image to determine the spot size. The spatial extent of the photocurrent is similar to the spot size for these room temperature measurements. This shows that the cooling length is relatively short, indicating that on a micron length-scale lateral heat transport is still slower than hot-carrier cooling (at room temperature).
}
\label{SuppFigSpatialImage}
   \end{figure}

\begin{figure} [h!!!!!]
   \centering
   \includegraphics [scale=0.8]
   {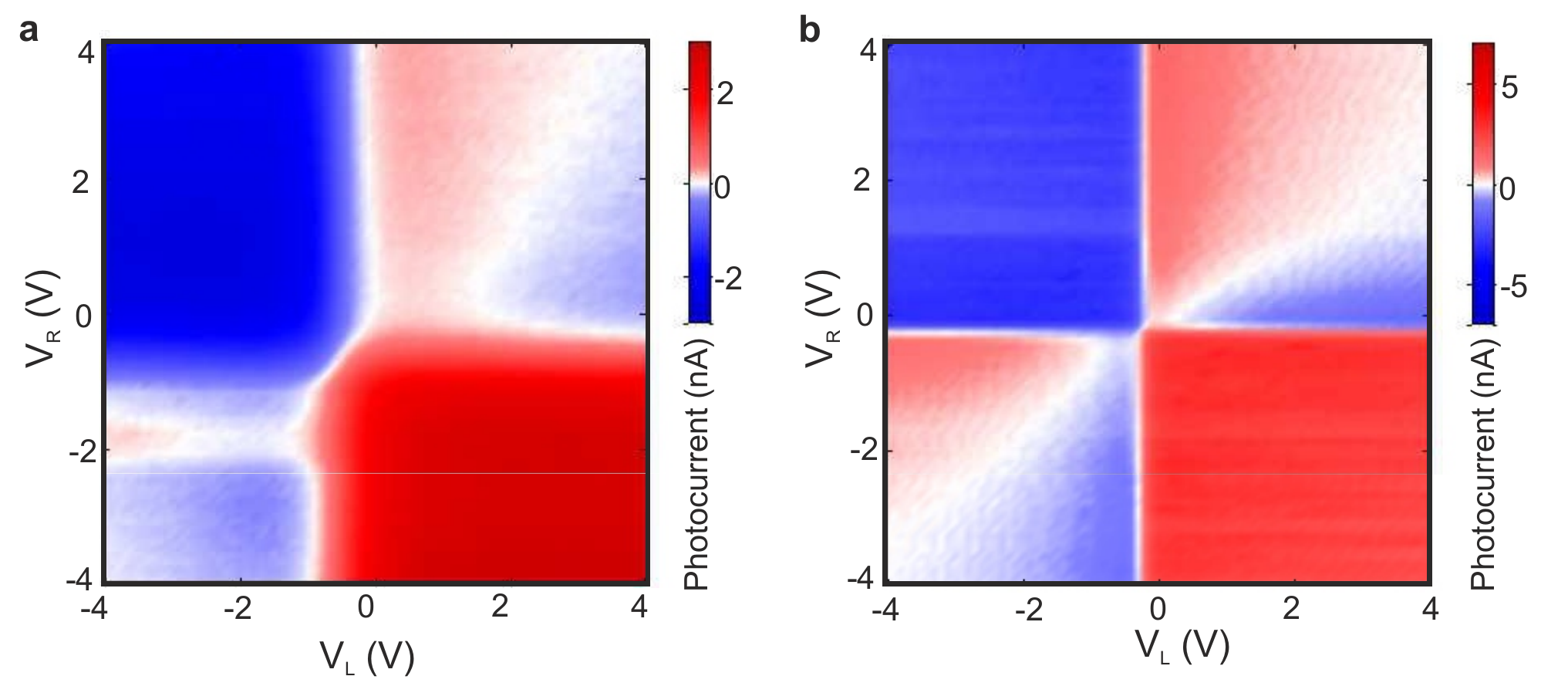}
\caption{\textbf{Characteristic sixfold pattern for photo-thermoelectric effect.} Photovoltage at the interface between the two graphene regions, for single-pulse excitation as a function of voltage on the two graphene split gates $V_L$ and $V_R$ at room temperature \textbf{(a)} and 30 K \textbf{(b)}. The sixfold patterns indicate photo-thermoelectric photovoltage generation, which scales with the light-induced increase in carrier temperature \cite{Gabor2011, Song2011}.
}
\label{SuppFigSixfold}
   \end{figure}

\begin{figure} [h!!!!!]
   \centering
   \includegraphics [scale=0.7]
   {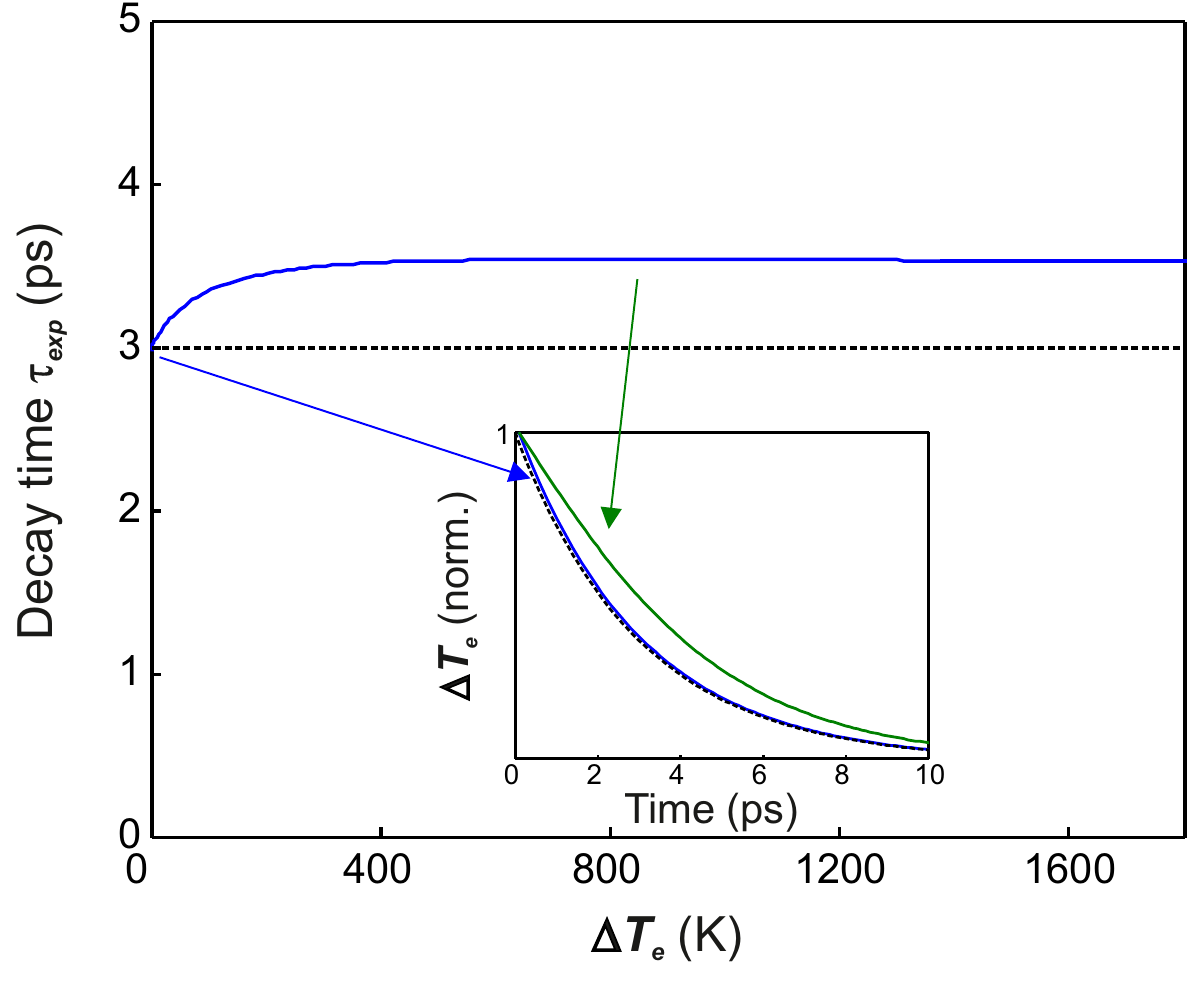}
     \caption{ \textbf{Effect of nonlinearity on measured dynamics. }
      To study the relation between the observed dynamics of the photovoltage dip and the actual cooling dynamics, we calculate the photovoltage dip dynamics for a given exponential cooling time of 3 ps. To this purpose, we integrate the generated photocurrent as a function of real time, after an ultrafast laser pulse heats up the carriers to a certain temperature that is related to the incident power and a second pulse heats up the carriers to a temperature that depends on the residual heat in the carrier system (following Ref.\ \cite{Graham2013}). The higher the power, the higher the peak temperature and the stronger the nonlinearity (due to the temperature-dependent electronic heat capacity). We find that the photocurrent dip as a function of delay time between the two ultrafast pulses gives an overestimation of the cooling dynamics by $<$20\%.  The inset shows the 'real' cooling dynamics (black dashed line), the photovoltage dynamics for low power (blue line) and for a laser power that corresponds to $T_{\rm e} \approx$ 1000 K (green line).
}
\label{SuppFigNonlinearArtefact}
   \end{figure}

To make a more quantitative comparison, we take into account that the measurements are typically done in the strong heating regime, where $T_{\rm e} \gg T_{\rm L}$. We first measure the cooling dynamics for increasing laser power and estimate the electron temperatures that correspond to each power from the characteristic power-dependent photoresponse (see \ref{SuppFigCarrierTemperature} and Methods), thus obtaining the exponential decay time $\tau_{\rm exp}$ vs. $T_{\rm e}$ (see \textcolor{blue}{\textbf{\ref{Fig4}a}}). In the experimentally accessible regime ($\Delta T_{\rm e} = T_{\rm e} - T_{\rm L} >$ 200 K) the $\tau_{\rm exp}$ increases with increasing $T_{\rm e}$ (measured for $n = 1.2\times 10^{12}$/cm$^2$). We compare this with the calculated cooling time $\tau_{\rm calc}$, which describes the "instantaneous" cooling time at a certain $T_{\rm e}$, and find quantitative agreement \textit{without any adjustable parameters}. The reason for the increasing cooling time with increasing $T_{\rm e}$ can be seen from Eq.\ (2) and by noting that $\mathcal{Q}/(T_{\rm e} - T_{\rm L})$ scales roughly linearly with $(T_{\rm e} - T_{\rm L})$. At the same time, the electronic heat capacity $C_n$, increases more than linearly with increasing $T_{\rm e}$ for large $T_{\rm e}$. This leads to a net increase in the cooling time with increasing $T_{\rm e}$. The agreement between experiment and calculation prompts us to make a more quantitative comparison for varying carrier density $n$. \textcolor{blue}{\textbf{\ref{Fig4}b}} shows $\tau_{\rm exp}$ for three different laser powers ($P =$ 21, 47 and 94 $\mu$W), corresponding to three different initial hot electron temperatures ($\Delta T_{\rm e} =$ 650, 950 and 1250 K), together with the calculated near-equilibrium cooling time $\tau_{\rm calc}^*$. The comparison between experimental results in the strong heating regime ($\tau_{\rm exp}$) and theoretical results close to equilibrium ($\tau_{\rm calc}^*$) is justified in \textcolor{blue}{\textbf{\ref{Fig4}a}}, where we show that the cooling time at $T_{\rm e} \sim$ 1000 K is similar to $\tau_{\rm calc}^*$. Again we find agreement between experiment and theory \textit{without any adjustable parameters}.
\

Finally, we compare the calculated time-domain cooling dynamics in the strong heating regime directly with the measured photocurrent dynamics. The calculated cooling dynamics describe cooling from an initial $T_{\rm e}$ down to $T_{\rm L}$, as constructed from the electron temperature-dependent cooling time $\tau_{\rm calc}$ in Eq. (2).  We compare these calculated cooling dynamics directly with the measured photovoltage dynamics $\Delta V_{\rm PTE} (\Delta t)$ for six different lattice temperatures, at a low carrier density of $n = 0.06 \times 10^{12}$/cm$^2$ (see \textcolor{blue}{\textbf{\ref{Fig4}c}}). Using the initial temperature (at the \textit{pn}-junction) as a fit parameter, we find that the hyperbolic cooling model can describe the experimental data very well. At room temperature, the initial temperature is similar to the one we calculate (see \ref{SuppFigCarrierTemperature}), whereas we find a lower initial temperature at lower lattice temperatures. This is due to the increased mechanical vibrations of the sample and lateral heat diffusion out of the laser spot at lower $T_{\rm L}$, which both lead to a larger photocurrent spot size and thus a lower effective initial $T_{\rm e}$ at the \textit{pn}-junction (see \ref{SuppFigHeatDiffusion}). For a lattice temperature of 50 K, we furthermore see an increasing discrepancy for a delay time $>$15 ps, which most likely indicates that hBN-phonons alone do not account for the complete cooling dynamics. Similarly, we find that at a higher carrier density of $n = 1.7 \times 10^{12}$/cm$^2$ the hyperbolic cooling model starts deviating from the experimental data already for $T_{\rm L} <$ 200 K (see \ref{SuppFigCoolingHighDoping}). Most likely, at sufficiently low lattice temperature and sufficiently high carrier density, cooling through optical phonons \cite{Kampfrath2005, Mihnev2016} and momentum-conserving cooling to acoustic graphene phonons \cite{Bistritzer2009, Ma2014, Jadidi2016} become the dominant channels.
\

In summary, we addressed the issue of out-of-plane heat transfer in the device architecture of hBN-encapsulated graphene. Combining experiments with microscopic theoretical calculations, we showed that the dominant cooling channel is the one in which heat transfer from hot carriers to the hBN polar substrate occurs via near-field coupling to hyperbolic phonon modes. This efficient mechanism explains the observation of a lower-than-expected cooling time in clean hBN-encapsulated graphene. We note that the observation of slower cooling for thin hBN encapsulation could have important implications for photodetection applications. In addition, we predict that significantly slower cooling can be achieved using alternative layered dielectrics, such as MoS$_2$. Furthermore, the near-field coupling we studied in this work between hot graphene carriers and hyperbolic hBN phonons may pave the way to novel approaches in fields such as nanophotonics, ultrahigh resolution microscopy and nanoscale thermal management.

\section*{Methods}

\textit{Extraction of the initial hot carrier temperature --- } The measured photovoltage $V_{\rm PTE}$ scales with the \textit{time-averaged} increase in electron temperature $\langle T_{\rm e} - T_{\rm L} \rangle$. In the case of cooling length smaller than the laser spot size (which is the case at room temperature) the relevant heat equation gives a simple linear scaling between peak and average increase in $T_{\rm e}$, governed by the interfacial heat conductivity $\Gamma$. We assume that $\Gamma$ and the Seebeck coefficient are constant with power \cite{Hwang2009a}, so that the photovoltage as a function of power directly represents the peak $T_{\rm e}$ increase as a function of power: $V_{\rm PTE} = a(T_{\rm e} - T_{\rm L})$. For undoped graphene, the peak $T_{\rm e}$ after illumination with laser power $P$ is $T_{\rm e} = \sqrt[3]{T_{\rm L}^3 + bP} $, where $b$ is a constant that depends on laser repetition rate $f$, absorption coefficient $\eta_{\rm abs}$, heating efficiency $\eta_{\rm heat}$ and spot size $L_{\rm spot}$. The cube root comes from the $T^2$-scaling of the electronic heat capacity for un-doped graphene (we apply this analysis to a low carrier density of $0.06 \times 10^{12}$/cm$^2$).  By fitting $V_{\rm PTE}$ as a function of $P$, we extract the constants $a$ and $b$, which allows us to recover the power-dependent peak $T_{\rm e}$ (see \ref{SuppFigHeatConductivity}). We verify the obtained constant $b$, and find good agreement when using an absorption of $\eta_{\rm abs}$ = 1\% (due to the layered dielectrics), a heating efficiency of $\eta_{\rm heat}$ = 80\% and a spot size of $L_{\rm spot} \sim$ 2 $\mu$m.\\

\textit{Numerical calculations ---} Theoretical curves are obtained by numerically integrating Eq.\ (1) \cite{Principi2016}.  The non-local conductivity is connected to the density-density response function of graphene, $\chi_{nn}(k,\omega)$, by the formula $\sigma(\omega,k) = i e^2\omega \chi_{nn}(k,\omega)/k^2$. Within the random-phase approximation, we approximate $\chi_{nn}(k,\omega)$ with its non-interacting expression, which has been given in Refs.\ \cite{Wunsch2006, Hwang2007, Principi2009, Kotov2012}. Neglecting retardation effects, the surface admittance is connected to the screened Coulomb interaction between two electrons in graphene, $V(k,\omega)$, by the formula $Y(\omega,k) = - i e^2 \omega V^{-1}(k,\omega)/k^2$. The expression for $V(k,\omega)$ is given, for graphene embedded into slabs of hBN and in the presence of non-hyperbolic dielectrics and metallic gates, in Ref.\ \cite{Gonzalez2017}.

\subsection*{Acknowledgments}
We thank Andrea Tomadin and Fabien Vialla for valuable discussions. This work was supported by the European Union's Horizon 2020 research and innovation programme under grant agreement No. 696656 Graphene Flagship, Fondazione Istituto Italiano di Tecnologia, the Spanish Ministry of Economy and Competitiveness through the Severo Ochoa Programme for Centres of Excellence in R\&D (SEV-2015-0522), Fundacio Cellex Barcelona, the  Mineco grants Ramon y Cajal (RYC-2012-12281), Plan Nacional (FIS2013-47161-P), and the Government of Catalonia trough the SGR grant (2014-SGR-1535), the ERC StG CarbonLight (307806), ERC Grant Hetero2D, EPSRC Grants EP/K01711X/1, EP/K017144/1, EP/N010345/1, and EP/L016087/1. K.J.T. acknowledges support through the Mineco Young Investigator Grant (FIS2014-59639-JIN). A.P. acknowledges support from the ERC Advanced Grant 338957 FEMTO/NANO and from the NWO via the Spinoza Prize. M.M. thanks the Natural Sciences and Engineering Research Council of Canada (PGSD3-426325-2012). D.T. acknowledges financial support from European Union Marie Curie Program (Career Integration Grant No. 334324 LIGHTER) and Max Planck Society. K.W. and T.T. acknowledge support from the Elemental Strategy Initiative conducted by the MEXT, Japan and JSPS KAKENHI Grant Numbers JP26248061, JP15K21722 and JP25106006.

\onecolumngrid

\section*{Appendix 1: Cooling channels for hot carriers in graphene}

Here we assess cooling channels for a hot-carrier distribution in graphene that compete with cooling through near-field coupling to hyperbolic hBN phonon polaritons. We mainly discuss supercollision and normal collision scattering to graphene acoustic phonons and optical graphene phonon cooling, while also briefly discussing flexural phonons, Wiedemann-Franz cooling and hot-carrier tunneling.
\

\subsubsection*{Cooling to graphene acoustic phonons}

\noindent \textit{Electron-acoustic phonon deformation potential} ---  Since coupling to graphene acoustic phonons (either through disorder-assisted supercollisions or through normal collisions) depends heavily on the electron-phonon deformation potential $D$, we first discuss its value. The value of $D$ is reasonably well established, with transport measurements on ultraclean, hBN-encapsulated devices giving $\sim$18--20 eV, assuming phonon-limited momentum scattering \cite{Dean2010, Wang2013}. The cooling dynamics of SiO$_2$-supported graphene are consistent with $D$ = 12--18 eV, assuming disorder-assisted cooling \cite{Graham2013}. We use transport measurements to determine the deformation potential for our device. Momentum-non-conserving collisions can occur because of: \textit{i)} long-range scattering mechanisms, \textit{ii)} short-range scattering mechanisms, and \textit{iii)} electron-phonon scattering \cite{DasSarma2011}. If momentum scattering is solely determined by electron-phonon interaction (process \textit{iii}), we can use the measured graphene mobility $\mu$ at a given carrier density $n$ to obtain the deformation potential following Refs.\ \cite{Hwang2008, Principi2014}: $\mu = \frac{4 \hbar v_{\rm F}^2 e \rho v_{\rm s}^2}{\pi D^2 n k_{\rm B} T}$, where $v_{\rm F}$ is the Fermi velocity, $\rho$ the mass density,  $v_{\rm s}$ the sound velocity and $\hbar$, $e$ and $k_{\rm B}$ the reduced Planck constant, electron charge and Boltzmann constant, respectively. Inserting relevant numbers gives $D =$ 35 eV. However, since the mean free path for low carrier concentrations scales linearly with $n$ (see \textcolor{blue}{\textbf{\ref{Fig2}a}}), we know that long-range scattering (process \textit{i}) also plays a role \cite{DasSarma2011} and therefore the value for $D$ is an \textit{upper} limit. Thus, we find that our transport data are consistent with $D <$ 35 eV.
\\

\noindent \textit{Supercollision cooling} --- The supercollision cooling mechanism relies on the presence of short-range scatterers (long-range scattering mechanisms give infinite cooling time) \cite{Song2012} and gives a cooling time that scales with $\frac{k_{\rm F}\ell}{D^2}$, where $k_{\rm F}$ is the Fermi momentum, $\ell$ the mean free path (limited by short-range scattering), and $D$ the electron-phonon deformation potential \cite{Graham2013}. We now proceed to calculate the deformation potential that would be necessary to reproduce the observed photovoltage dip, if supercollision scattering would dominate, where we closely follow Ref.\ \cite{Graham2013} in calculating the experimental photovoltage dip $\Delta V_{\rm PTE}$. Since this model requires that $T_\mathrm{el} \ll T_\mathrm{F}$, we apply this analysis to our data at $n = 1.7 \times 10^{12}$/cm$^2$, where $T_F = \frac{E_F}{k_{\rm B}} \approx 1800$ K.  We numerically solve the energy dissipation rate $C_n \frac{\mathrm{d} T_{\rm e}}{\mathrm{d}t} = -A(T_{\rm e}^3-T_{\rm L}^3)$ with the heat capacity (for non-neutral graphene) $C_n = \alpha T_\mathrm{el}$ and $\frac{A}{\alpha} =  0.47 \frac{1}{k_F \ell}  \frac{D^2}{\rho v_{\rm s}^2}  \frac{ E_F}{(\hbar v_F)^2 )} \frac{k_{\rm B}}{\hbar} $ to obtain the temperature dynamics $T_{\rm e}(t)$. The photovoltage $V_{\rm PTE}$ at a delay time $\Delta t$ between two pulses follows from $V_{\rm PTE} = \int_0^{\Delta t} V(t, T_\mathrm{1}) \mathrm{d}t  + \int_{\Delta t}^{\infty} V(t-\Delta t, T_\mathrm{2}) \mathrm{d}t$. Here we use the instantaneous photo-thermoelectic voltage $V (t, T) = BT(T - T_{\rm L})$, where $B$ is a proportionality constant related to the Seebeck effect. The initial hot-electron temperature after the first (second) laser pulse $T_\mathrm{1}$ ($T_\mathrm{2}$) is given by $\sqrt{T_\mathrm{in}^2+T_\mathrm{add}^2}$, with $T_\mathrm{in}$ the temperature before arrival of the laser pulse, and $T_\mathrm{add}$ the temperature equivalent to the added pulse energy. We obtain a photovoltage dip by repeating this calculation while varying $\Delta t$, and fit the data to extract the deformation potential, finding $D$ = 65 eV (see \textcolor{blue}{\textbf{\ref{SuppFigSupercollision}a}}). We note that using Ref.\ \cite{Song2012} to relate $\frac{A}{\alpha}$ to $D$ would give a $D$ that is a factor $\sqrt{8}$ higher. Thus, we find that in order to reproduce the observed cooling dynamics, the deformation potential would have to be an unrealistically high $D >$ 65 eV. This value is a \textit{lower} bound, because the analysis assumes that all scatterers that lead to the mean free path also contribute to supercollision scattering, whereas actually only a fraction -- the short-range scatterers -- contribute \cite{Song2012}.  Since transport measurements indicate $D <$ 35 eV and cooling dynamics indicate $D >$ 65 eV, we conclude from this quantitative analysis that supercollision is likely not the dominant cooling mechanism in hBN-encapsulated graphene. This is corroborated by the observed trend that cooling becomes faster with increasing carrier density -- exactly opposite to the trend that is measured for supercollision-dominated devices where faster cooling is observed around the Dirac point \cite{Graham2013, Graham2013a}. Finally, noise thermometry studies on hBN-encapsulated graphene \cite{Crossno2015} and experimental-theoretical terahertz spectroscopy studies on multilayer epitaxial graphene and CVD graphene on different substrates\cite{Mihnev2016} both led to the conclusion that supercollision cooling does not explain the experimental results.
\\

\noindent \textit{Normal collision cooling} --- Cooling to acoustic graphene phonons without disorder-assisted scattering is generally believed to be slow, with typically nanosecond timescales \cite{Bistritzer2009}. However, in the regime where the electron temperature exceeds the lattice and Fermi temperatures, cooling occurs significantly faster. In this regime, cooling is governed by non-exponential cooling dynamics, according to \cite{Bistritzer2009} $T_{\rm e} (\Delta t) = \frac{T_{\rm e} (0)}{\sqrt{\Delta t/\tau_0 + 1}}$, with $\tau_0 = \frac{424}{D^2 [T_{\rm e} (0)]^2}$, where $D$ is in eV, $T_{\rm e}$ in meV and $\tau_0$ in $\mu$s. We compare these cooling dynamics with our experimental data in \textcolor{blue}{\textbf{\ref{SuppFigSupercollision}b}}, showing that our experimental decay is significantly faster. For the calculated dynamics we use $D$ = 35 eV, the maximum value that is consistent with our transport data. Using lower (more realistic) values for $D$, cooling would be even slower. We also compare the calculated dynamics  with exponential decay with a cooling time of 28 ps, which corresponds to the fastest (initial) decay of these cooling dynamics. This is roughly one order of magnitude slower than the experimentally observed dynamics. From this quantitative analysis we conclude that cooling to acoustic graphene phonons is likely not the dominant cooling mechanism. Qualitatively, we observe exponential decay dynamics (at least above a lattice temperature of 200 K) instead of the predicted non-exponential cooling. Furthermore, cooling through normal collisions with graphene acoustic phonons in the overheating regime is predicted to be independent of lattice temperature, whereas we observe clearly slower cooling at lower lattice temperatures.
\\

\subsubsection*{Cooling to graphene optical phonons}

\begin{figure} [h!!!!!]
   \centering
\includegraphics [scale=0.8]
   {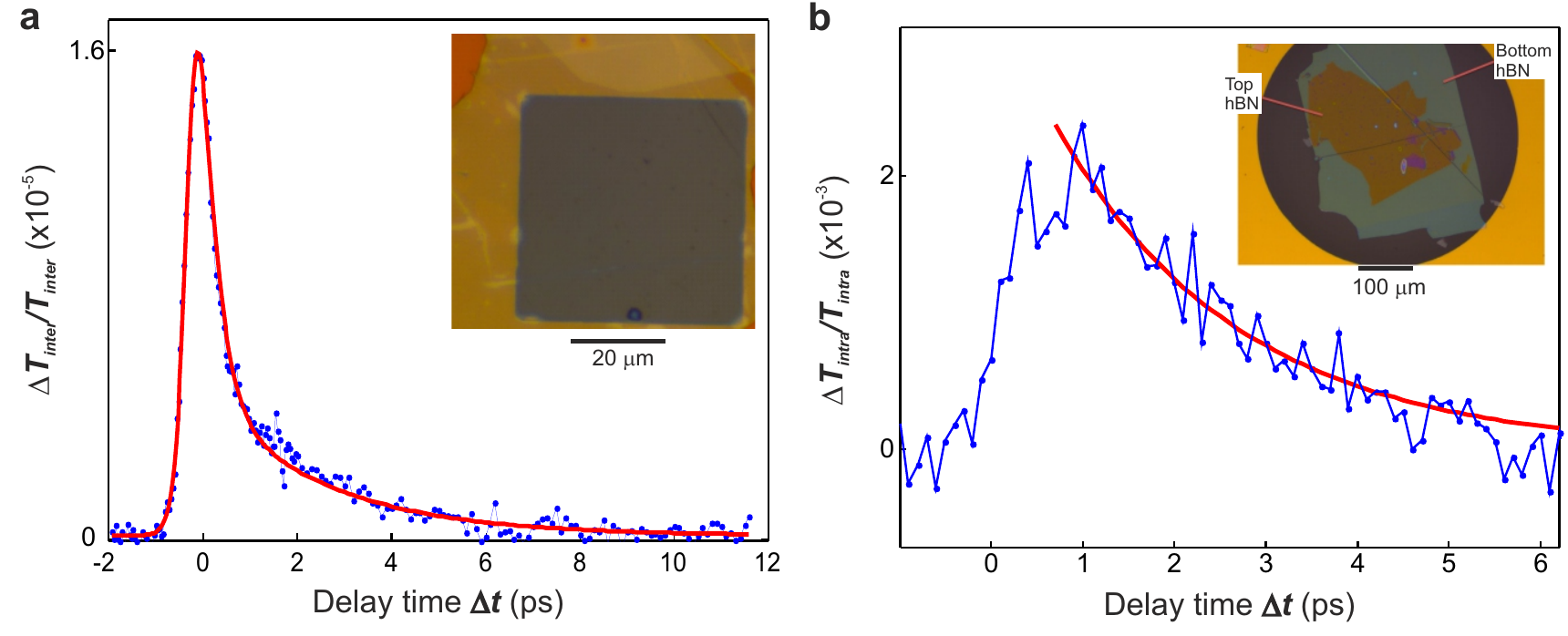}
\caption{ \textbf{Verification of RT decay time using pump-probe spectroscopies.}
\textbf{a)} Ultrafast optical pump -- optical probe spectroscopy on hBN-encapsulated CVD graphene, prepared as in Ref.\ \cite{Banszerus2015} and fully covering a 50 $\mu$m square aperture on a transparent SiO$_2$ substrate (see inset). The ultrashort pump pulses, with a wavelength of 780 nm, 100 fs-duration, and incident fluence of $\sim$10 $\mu$J/cm$^2$, creates a non-equilibrium distribution that quickly thermalizes through electron-electron scattering \cite{Brida2013}. This modified, hot-carrier distribution affects available interband transitions through Pauli blocking, which we probe using ultrashort pulses at 1.3 $\mu$m. The differential transmission $\Delta T_{\rm inter}/T_{\rm inter}$ presents a bi-exponential decay, with the slower component representing the main hot-electron cooling channel with a time scale of $\sim$2.5 ps. The fast initial decay is likely related to the super-linear relation between changes in the electron temperature and changes in the differential transmission. This fast decay component is not seen in time-resolved photocurrent and optical pump -- terahertz probe measurements since these techniques both have a sub-linear relation between changes in the electron temperature and changes in the observed signal (see also \ref{SuppFigCarrierTemperature}). 
\textbf{b)} The pump-induced change in transmitted terahertz (THz) signal as a function of pump-probe delay time $\Delta t$ using an hBN-encapsulated CVD graphene sandwich with dimensions $>$200 $\mu$m, prepared as in Ref.\ \cite{Banszerus2015} and transferred inside a 500 $\mu$m round aperture (see inset). The loosely focused pump light, with a wavelength of 800 nm and pulse energy density of $\sim$ 20 $\mu$J/cm$^2$ creates hot carriers and a modified carrier distribution. Terahertz light with frequency 0.4 -- 1.2 THz subsequently probes the effect of the modified distribution on intraband transitions, i.e.\ the electronic response. The pump-probe dynamics of the differential THz transmission $\Delta T_{\rm intra}/T_{\rm intra}$ represent hot-carrier cooling (for carriers with energy below the optical phonon energy) and give a timescale of $\sim$2.2 ps.
}
\label{SuppFigPumpProbe}
   \end{figure}

\begin{figure} [h!!!!!]
   \centering
\includegraphics [scale=0.7]
   {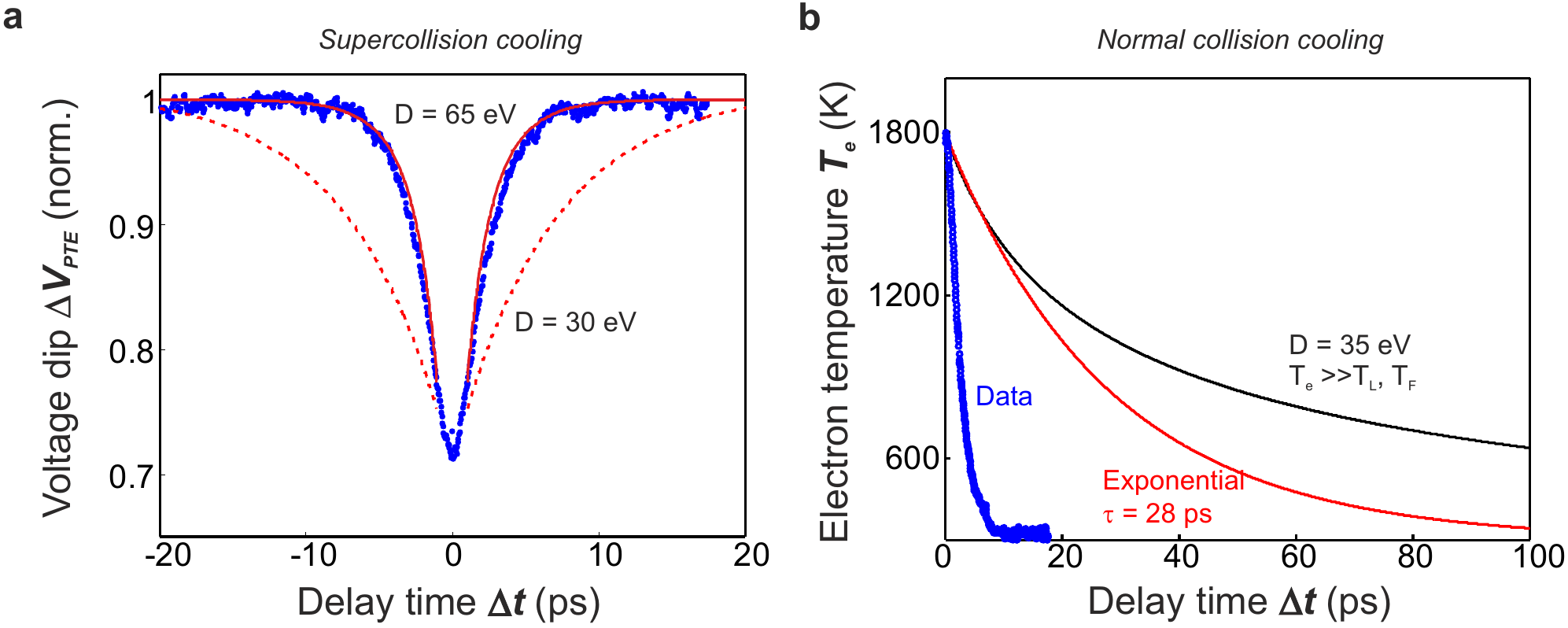}
\caption{ \textbf{Comparing data with alternative cooling mechanisms.}
The photovoltage dip dynamics ($T_{\rm L}$ = 300 K and $n = 1.7\times 10^{12}$/cm$^2$, compared with the dynamics according to the supercollision cooling model with a deformation potential of 65 eV (solid red line) and 30 eV (dashed red line). The model takes into account that we are in the strong heating regime with $T_{\rm e} = 1500$ K (see Methods).
}
\label{SuppFigSupercollision}
   \end{figure}

\begin{figure} [h!!!!!]
   \centering
   \includegraphics [scale=0.7]
   {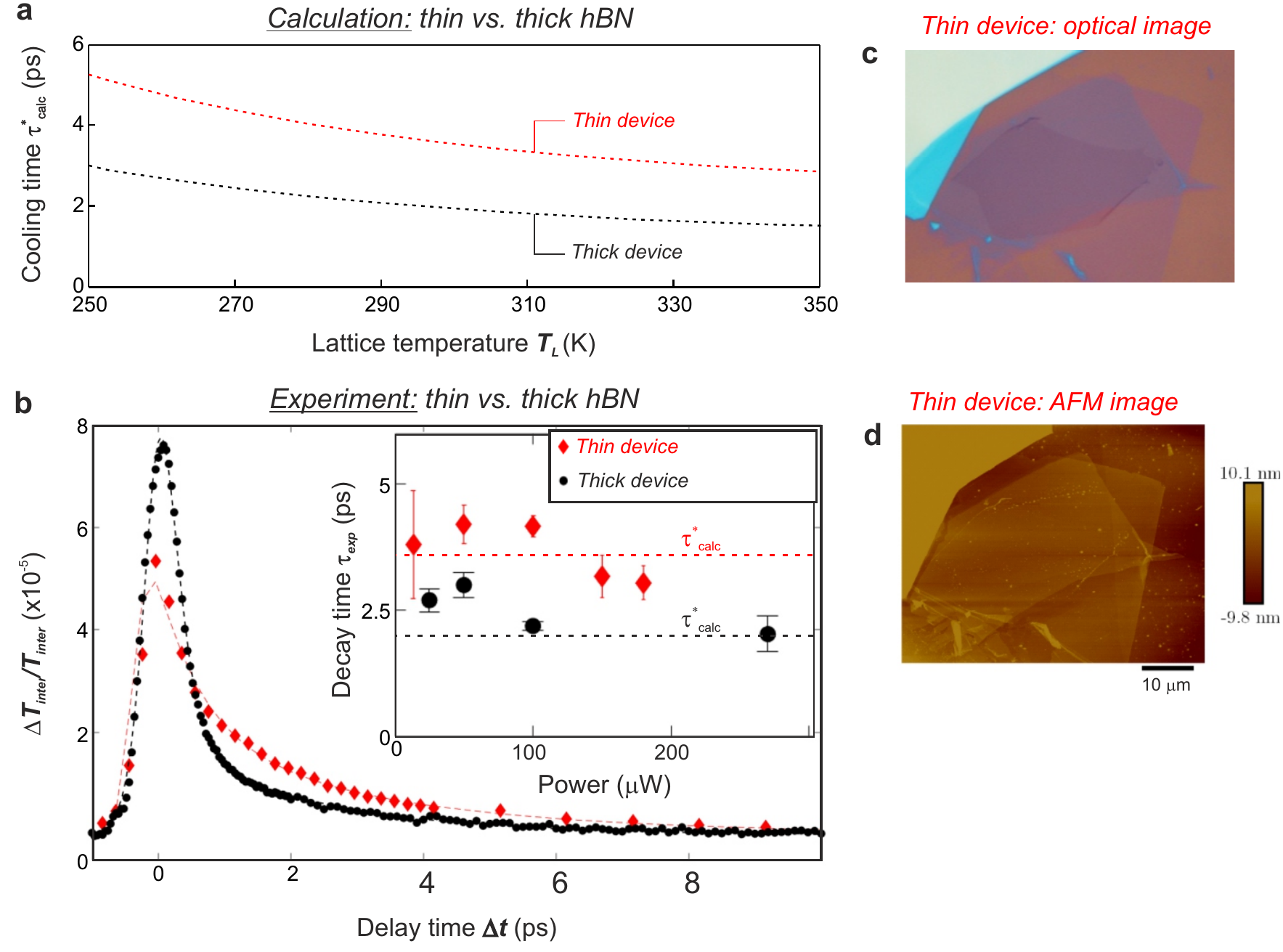}
\caption{\textbf{Dynamics for different hBN thicknesses.}
\textbf{a)} Calculated (near-equilibrium) cooling time according to the hyperbolic cooling model $\tau_{\rm calc}^*$ for a "thick" device with 30 nm bottom and 30 nm top hBN encapsulant (black, dashed line) and a "thin" device with 2 nm top and 3 nm bottom hBN encapsulant (red, dashed line). Clearly, cooling is slower in the device with very thin encapsulation.
\textbf{b)} Differential transmission as a function of pump-probe delay time for graphene/hBN devices with different encapsulant thickness, as in panel \textbf{b}. For both devices, the probe beam was around 1300-1400 nm. The wavelength of the pump beam was 1550 nm (785 nm) for the “thin” (“thick”) device and the power 180 (100) $\mu$W. We compare measurements taken at slightly different powers, because due to the difference in focus spot size, these powers result in similar initial carrier temperatures for both measurements (hence the similar size of the peak differential transmission right after time zero).  We point out that the pump wavelength is not relevant for the picosecond cooling dynamics, since within tens of femtoseconds carrier heating occurs and any memory of the initial energy of the photoexcited e-h pairs is lost.
The inset of panel \textbf{b} gives the experimental decay times, which are the slower time constants from bi-exponential fits, corresponding to carrier cooling. We compare these experimental decay times with the calculated cooling times (dashed lines). Panels \textbf{c)} and \textbf{d)} give optical and AFM images of the “thin” device. We note that both devices have similar Fermi energy as extracted from Raman measurements (not shown), with the "thick" ("thin") device having a G peak at 1584 (1583) cm$^{-1}$ and a 2D peak at 2686 (2685) cm$^{-1}$. 
}
\label{SuppFigThinHBN}
   \end{figure}

\begin{figure} [h!!!!!]
   \centering
\includegraphics [scale=0.6]
   {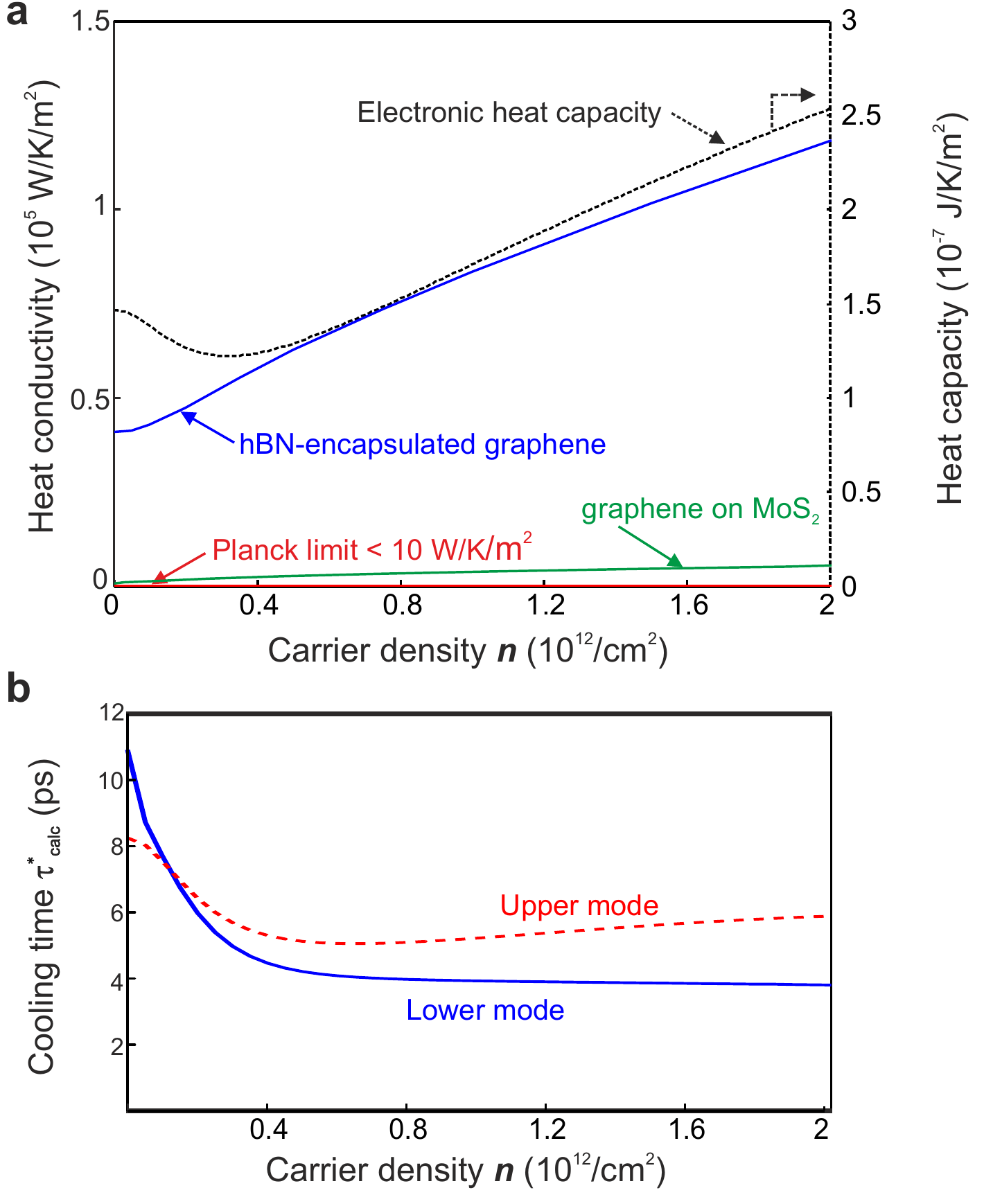}
\caption{\textbf{Interfacial heat conductivity, heat capacity and contributions of the hyperbolic modes.}
\textbf{a)} Since the hyperbolic cooling time $\tau_{\rm calc}^*$ is determined by the ratio of the interfacial heat conductivity $\Gamma$ and the electronic heat capacity $C_n$, we plot these separate entities as a function of carrier density $n$. We also show the interfacial heat conductivity for MoS$_2$ hyperbolic phonons, and for black-body radiation of graphene in vacuum, which is orders of magnitude lower than for hBN or MoS$_2$.
\textbf{b)} The separate contributions to the cooling time from the two distinct hyperbolic phonon polariton modes in hBN: the 'Lower' mode at 100 meV and the 'Upper' mode at 180 meV.
}

\label{SuppFigHeatConductivity}
   \end{figure}

\begin{figure} [h!!!!!]
   \centering
   \includegraphics [scale=0.6]
   {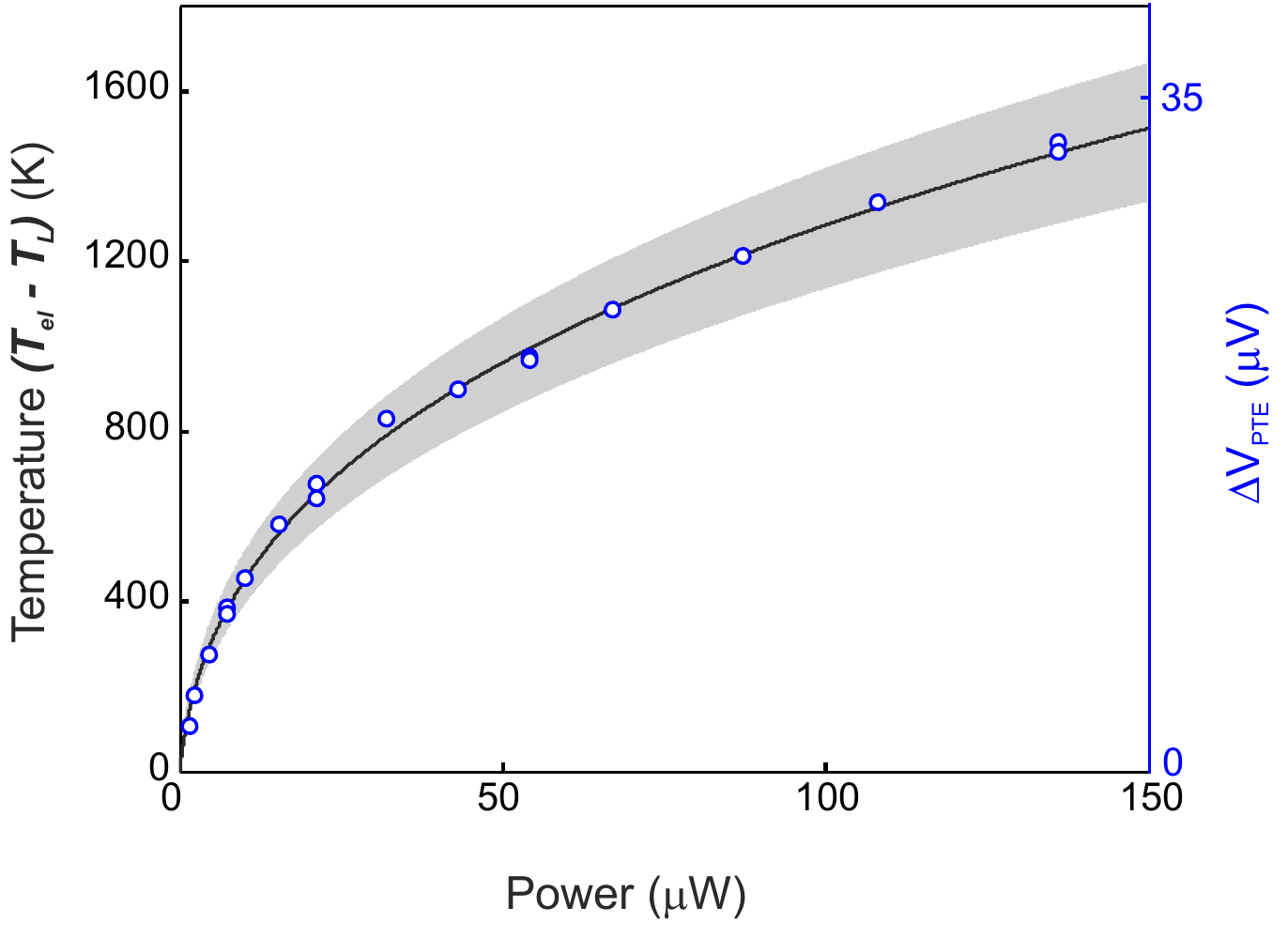}
\caption{ \textbf{Estimation of hot carrier temperature. }
      The photovoltage as a function of laser power (right vertical axis) and the extracted laser-induced temperature increase (left vertical axis) for undoped graphene, using the analysis described in the Methods section. The grey shaded area gives the 95\% confidence interval for the extracted electron temperature. 
 }
\label{SuppFigCarrierTemperature}
   \end{figure}

\noindent There is general consensus in the literature that cooling to \textit{graphene optical phonons} occurs on a rapid (sub-picosecond) timescale \cite{Kampfrath2005,Brida2013}, but mainly for carriers with high energy, on the order of the optical phonon energy of $\sim$0.2 eV \cite{Viljas2010}. Therefore, one would expect this process to mainly play a role at sufficiently high fluences. On the other hand, a recent study on \textit{non-encapsulated graphene devices} showed that even at moderate fluences (similar to the ones used in our study) cooling by coupling to optical phonons plays a role for the observed picosecond decay dynamics of the THz photoconductivity, as measured by optical pump -- THz probe spectroscopy (the same technique that we used in  {\textcolor{blue}{\textbf{\ref{SuppFigPumpProbe}b}}}). This prompts us to study in more detail if cooling to optical phonons could explain our experimental results. Comparing our data and the results in Ref.\ \cite{Mihnev2016}, we see that some trends are similar for cooling to graphene optical phonons and for cooling to hBN hyperbolic phonons. In particular, both cooling mechanisms lead to slower cooling for decreasing Fermi energies and for increasing fluence. However, there are also a number of distinct differences: \textit{i)} We observe significantly slower cooling for lower substrate temperatures (for all examined equilibrium Fermi energies, $E_{\rm F}$ = 0.04 –- 0.17 eV), e.g.\ a twofold increase in cooling time by decreasing $T_{\rm L}$ from 300 K to 200 K, whereas optical phonon cooling predicts that the cooling dynamics are independent of substrate temperature down to 50 K or less (for $E_{\rm F}$ = 0.3 eV) \cite{Mihnev2016}; \textit{ii)} The optical phonon cooling model predicts either bi-exponential cooling dynamics for $E_{\rm F}$ = 0.3 eV, or rather slow cooling with a time constant of 5-7 ps at $E_{\rm F} \approx$ 0 eV (both at room temperature) \cite{Mihnev2016}, which is not consistent with our observations at room temperature, where we see exponential decay with a decay constant well below 4 ps. Furthermore, we point out that a recent noise thermometry study also showed that cooling through optical phonons plays a minor role for hBN-encapsulated graphene \cite{Yang2017}. Finally, we observe a significant effect of reducing the thickness of the hBN encapsulant (see { \ref{SuppFigThinHBN}}), which indicates that the observed picosecond cooling dynamics are related to the encapsulation material. Thus, whereas a fraction of the energy of hot electrons is likely lost on a sub-picosecond timescale to optical phonons, the main cooling channel giving rise to the observed picosecond cooling dynamics corresponds to hyperbolic phonon cooling. 
\\

\subsubsection*{Other cooling channels}

\begin{figure} [h!!!!!]
   \centering
   \includegraphics [scale=0.7]
   {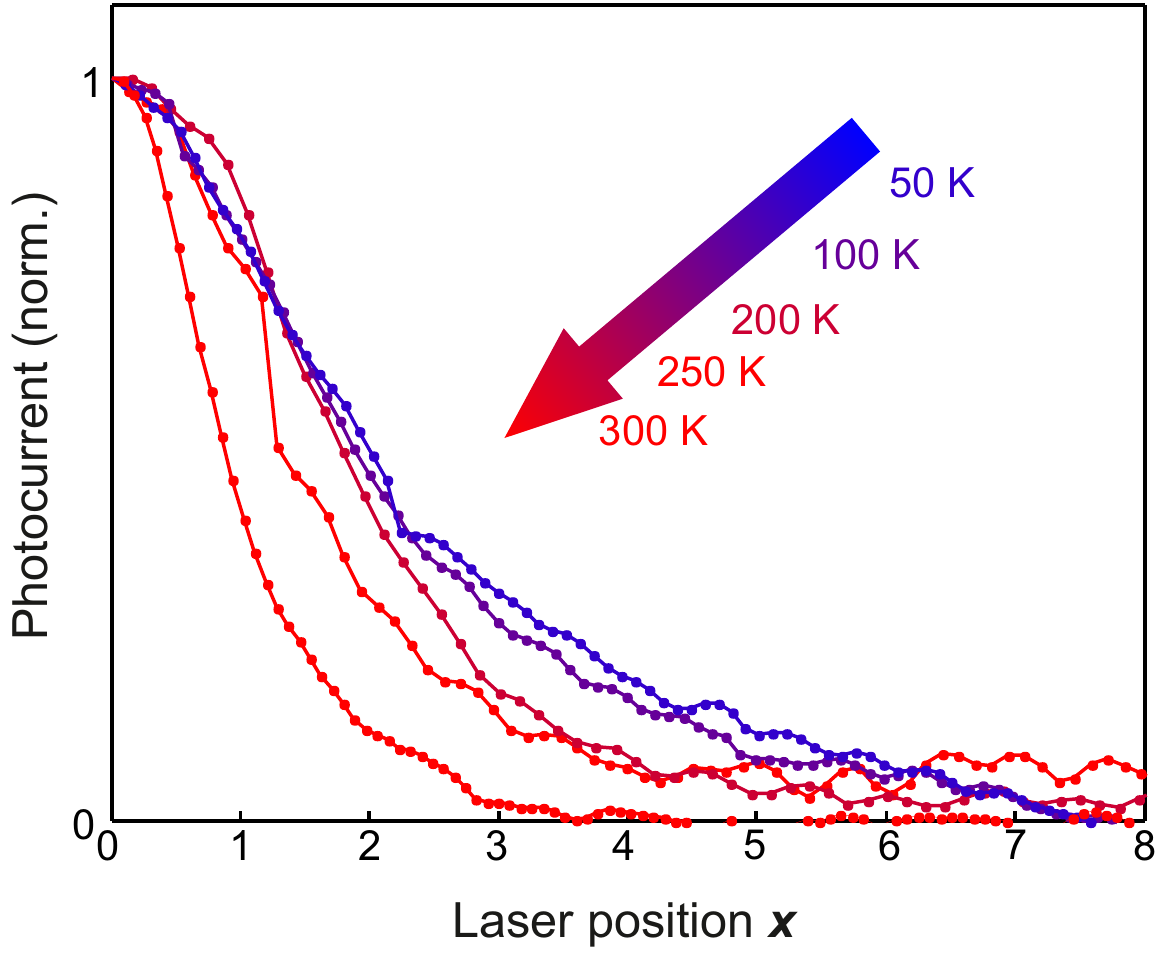}
      \caption{ \textbf{Spatial extent of photocurrent.}
      Spatial line traces of photocurrent vs. laser position for five different lattice temperatures. Position 0 corresponds to the $pn-$junction. At lower temperatures, the spatial extent of the photocurrent increases. This is caused by increased mechanical vibrations of the sample (due to the compressor of our helium closed cycle cryostat) and by a longer cooling length (due to slower hyperbolic hot-carrier cooling and a longer mean free path). This means that lateral heat diffusion out of the laser spot leads to an additional cooling channel, i.e.\ Wiedemann-Franz cooling. As a result of the larger photocurrent spot size the initial temperature is significantly lower for lower lattice temperature.
}
\label{SuppFigHeatDiffusion}
   \end{figure}

\begin{figure} [h!!!!!]
   \centering
  \includegraphics [scale=0.7]
   {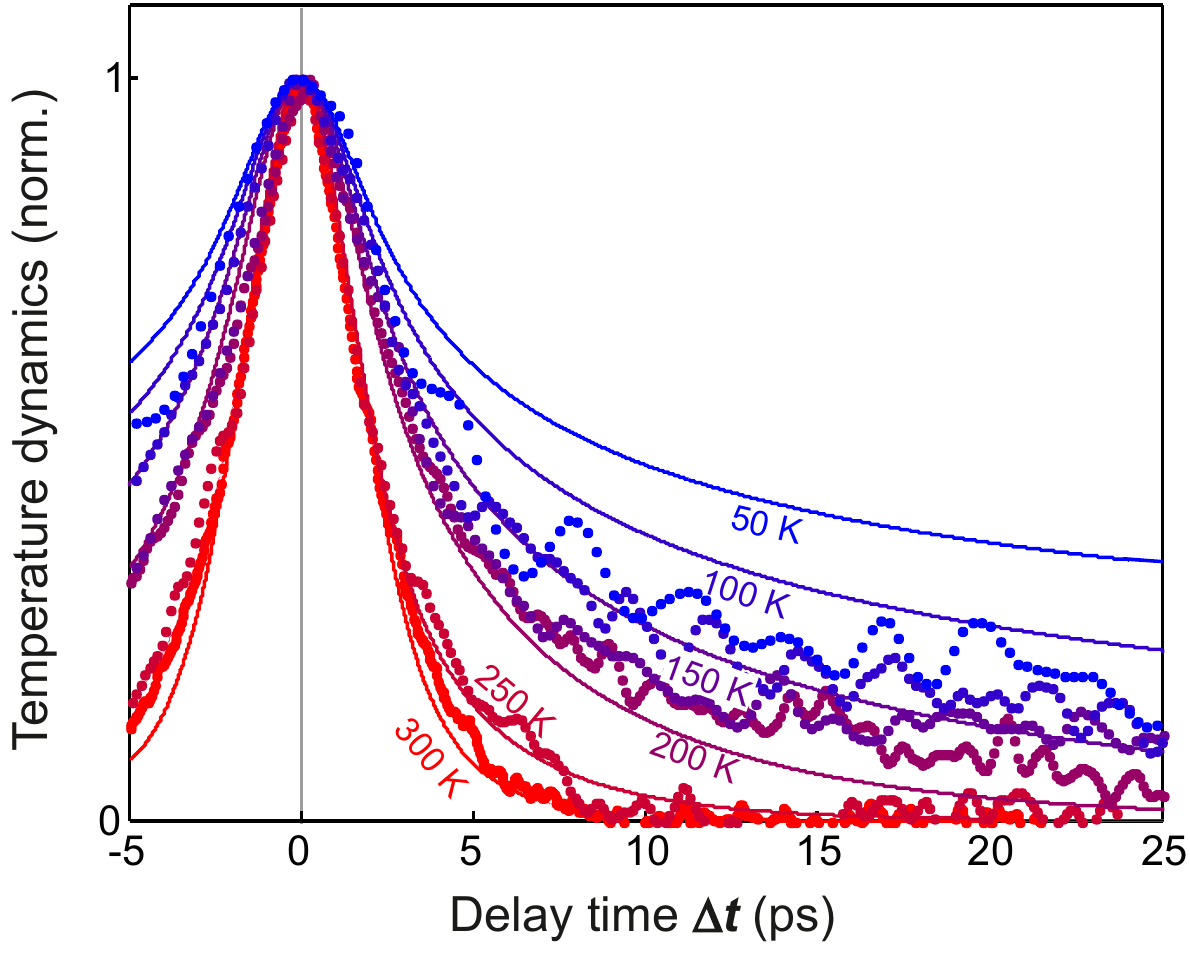}
      \caption{ \textbf{Cooling dynamics at high doping.}
Comparison of the complete cooling dynamics as measured (data points) and calculated (solid line) for  graphene with $n$ = 1.7 and 2 $\times 10^{12}$/cm$^2$, respectively, and varying lattice temperature. We use an initial temperature increase of $\Delta T =$ 300--500 K for all temperatures, except 1250 K for $T_{\rm L}$ = 300 K, the same as we did for undoped graphene. We see deviations between data and model starting from $T_{\rm L}$ = 150 K, most likely caused by normal collision with graphene acoustic phonons \cite{Bistritzer2009} and scattering with optical phonons \cite{Mihnev2016} becoming more relevant cooling pathways.
}
\label{SuppFigCoolingHighDoping}
   \end{figure}

\noindent Concerning \textit{flexural phonons}, we expect that these will give an even smaller contribution than normal collision scattering with graphene acoustic phonons, because they are quenched by the encapsulation. Even taking into account the elasticity of the substrate, the main branch of the flexural modes is predicted to be (i) gapped and (ii) strongly damped \cite{Amorim2013}. \textit{Wiedemann-Franz cooling} refers to lateral heat spreading out of the spot that is exited by the incident light. To assess the relevance of this process, we calculate the cooling length (at room temperature), which is given by $\zeta = \sqrt{\frac{\kappa}{\Gamma}}$ \cite{Song2011}. Here, $\Gamma$ is the interfacial heat conductivity and $\kappa$ is given by the Wiedemann-Franz law: $\kappa = \frac{\pi^2k_{\rm B}^2T\sigma}{e^2}$, where $\sigma$ is the graphene electrical conductivity and $e$ the electron charge. We find a cooling length of $\zeta \sim$ 1 $\mu$m, which is smaller than our spot size of $L_{\rm spot} \sim$ 2 $\mu$m (and much smaller than the device size), and thus this lateral heat spreading process is irrelevant for our results at room temperature. We note that at lower temperatures, the conductivity increases and the interfacial heat conductivity $\Gamma$ decreases, leading to longer cooling lengths. Therefore, at lower lattice temperatures, we do not exclude that Wiedeman-Franz cooling plays a role (see {\ref{SuppFigHeatDiffusion}}). Finally, we mention that \textit{cooling by tunneling of hot carriers through the hBN slab} is irrelevant for our devices, since the hBN slab in between the graphene and the bottom gate has a thickness of 70 nm, whereas tunnelling only plays a role when the two graphene sheets are separated by hBN of just a few layers thick \cite{Britnell2012}.
\\

\section*{Appendix 2: Fabrication of thin hBN encapsulated sample}

Hexagonal BN and graphene flakes are first produced by micro-mechanical cleavage from bulk crystals onto silicon substrates coated with $285\text{nm}$ silicon dioxide (SiO$_{2}$). Suitable hBN and single layer graphene flakes for the encapsulation are then identified by bright and dark field optical microscopy \cite{Gorbachev2011,CasiNL7} and optical contrast measurements \cite{CasiNL7} and Raman spectroscopy \cite{PRL2006,NN2013,Arenal}. The target flakes are then picked up and assembled into the desired heterostructure using a hot-pick up technique, similar to Ref.\ \cite{Pizzocchero2016}, to minimize contamination containing blisters at the hBN/graphene interfaces. From atomic force microscopy (AFM) characterization (see \textcolor{blue}{\textbf{\ref{SuppFigThinHBN}d}}) we measure a step change in height$\sim$ 2\text{nm} and $\sim 3\text{nm}$ for the bottom and top hBN respectively. We note that part of this contribution (up to $\sim1\text{nm}$) may arise from a water layer present on the surface of the hBN \cite{Pizzocchero2016}, so that we estimate up to 7 layers in hBN encapsulants.
\

\clearpage
\section*{Appendix 3: Additional figures}

\end{document}